# Guided mode evolution and ionization injection in meter-scale multi-GeV laser wakefield accelerators

J.E. Shrock[1], E. Rockafellow[1], B. Miao[1], M. Le[1], R.C. Hollinger[2], S. Wang[2], A.J. Gonsalves[3], A. Picksley[3], J.J. Rocca[2,4], and H.M. Milchberg[1,5]

[1]Institute for Research in Electronics and Applied Physics and Department of Physics, University of Maryland, College Park, Maryland 20742, USA
[2]Department of Electrical and Computer Engineering, Colorado State University, Fort Collins, Colorado 80523, USA
[3]Lawrence Berkeley National Laboratory, Berkeley, California 94720, USA
[4]Department of Physics, Colorado State University, Fort Collins, Colorado 80523, USA
[5]Department of Electrical and Computer Engineering, University of Maryland, College Park, Maryland 20742, USA

We show that laser wakefield electron accelerators in meter-scale, low density hydrodynamic plasma waveguides operate in a new nonlinear propagation regime where sustained beating of lowest order modes of the ponderomotively modified channel plays a significant role, whether or not the injected pulse is linearly matched to the guide. For a continuously doped gas jet, this mode beating effect leads to ionization injection and a striated multi-GeV energy spectrum of multiple quasi-monoenergetic peaks; the same process in a locally doped jet produces single multi-GeV peaks with <10% energy spread. A 3-stage model of drive laser pulse evolution and ionization injection characterizes the beating effect and explains our experimental results.

Laser-driven plasma wakefields [1,2] have accelerating gradients ~1000 times larger than those in linear RF accelerators, enabling the production of multi-GeV electron bunches over just centimeters [3-6]. In general, optical guiding is needed to keep the intensity of the driving laser high and extend the acceleration length. While relativistic self-guiding has resulted in ~2 GeV electron bunches, it demands petawatt-scale laser powers [3,4]. More laser-energy-efficient and controllable acceleration of electrons can be achieved by using plasma waveguides [5-22].

Recently, we demonstrated laser wakefield acceleration (LWFA) of electrons to 5 GeV in a 20 cm plasma waveguide using <300 TW of laser power [6]. In our scheme, which employs hydrodynamic plasma waveguides [7,8] generated in Bessel-beam-heated meter-scale gas jets [6, 22], the laser pulse undergoes "self-waveguiding" [20,21], while simultaneously driving the plasma wake responsible for electron acceleration. First, a zeroth order ultrashort pulse Bessel beam, whose focus extends well beyond 20 cm, ionizes and heats (by optical-field ionization (OFI) [12-22]) a 20 cm long hydrogen gas sheet. The explosive hydrodynamic response leaves a $\sim 10 \times$ reduced gas density on axis surrounded by an expanding cylindrical shock shell of enhanced density measured by 2-color interferometry [20]. When end-injected by a high intensity laser pulse, the core and inside walls of this "prepared index structure" are ionized by the very early leading edge of the injected pulse, forming an on-the-fly plasma waveguide that confines the vast bulk of the pulse. The accelerated electrons are injected into the wake by ionization [23-26] from 5% nitrogen dopant gas.

In this paper, we present experiments and simulations demonstrating that laser intensity oscillations from a general mode beating effect --*active whether or not the injected laser is linearly* *mode matched to the guide*-- are responsible for repeated ionization injection, giving rise to characteristic multi-peaked electron spectra at multi-GeV energies. A 3-stage model of intense laser propagation explains our results. Insight from this model then motivates an experiment showing that mode beating at an axially localized dopant region produces quasi-monoenergetic multi-GeV electron bunches with energy spread <10% and mrad divergence. The mode beating-induced injection effect described in this work is new and enhanced by narrow, low density meter-scale plasma waveguides.

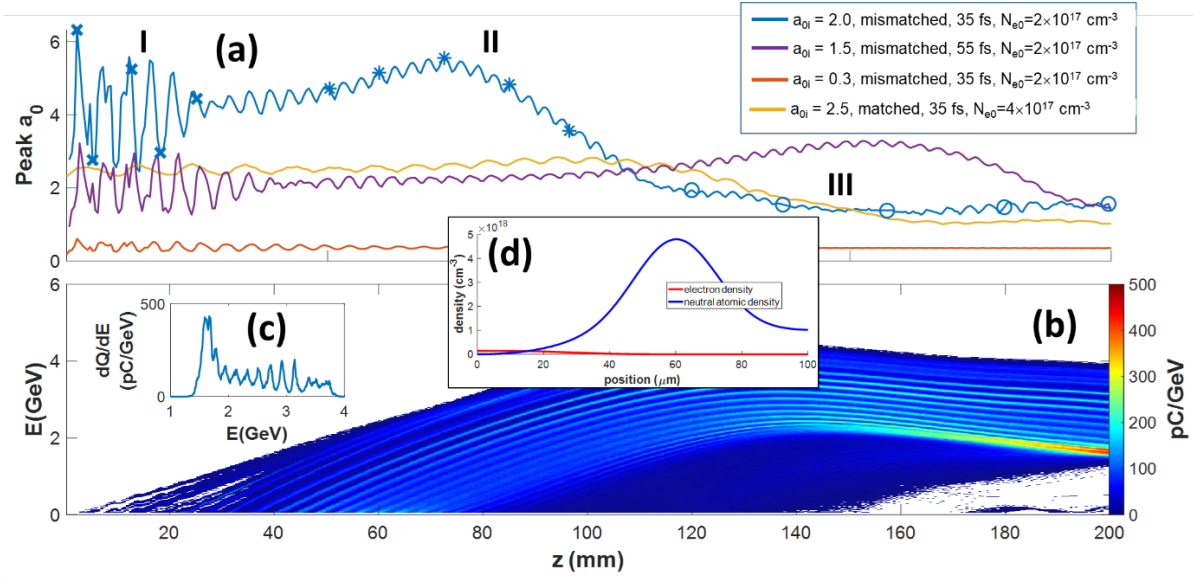

**Figure 1.** WarpX [27] particle-in-cell simulations of drive pulse evolution in a waveguide formed from the prepared index structure [20] shown in the central inset, with $w_{ch} = 20$ μm. Gas composition is 95%/5% $H_2/N_2$ and the simulation $xyz$ grid is 256×256×4096 (256μm×256μm×204.8μm). **(a)** Peak laser field $a_0$ vs. propagation distance for: (*i*) mismatched input field $a_{0i} = 2.0, w_0 = 30$ μm, $\tau_{fwhm} = 35\,fs$, $N_{e0} = 2.0 \times 10^{17}$ cm$^{-3}$ (blue curve), (*ii*) mismatched input field $a_{0i} = 1.5, w_0 = 30$ μm, $\tau_{fwhm} = 55\,fs$, $N_{e0} = 2.0 \times 10^{17}$ cm$^{-3}$ (purple curve) (*iii*) mismatched $a_{0i} = 0.3, w_0 = 30$ μm, $\tau_{fwhm} = 35$ fs, $N_{e0} = 2.0 \times 10^{17}$ cm$^{-3}$ (orange curve), and (*iv*) matched $a_{0i} = 2.5, w_0 = 30$ μm, $\tau_{fwhm} = 35\,fs$, $w_{ch} = 30\,\mu m$, $N_{e0} = 4.0 \times 10^{17}$ cm$^{-3}$ (gold curve, Gaussian input pulse, parabolic channel). Labels I, II, and III denote the three characteristic propagation stages. Labels (×), (∗), and (o) denote the locations of the frames displayed in Fig. 2(a)-(c). **(b)** Electron spectrum vs. propagation distance corresponding to the blue curve of (a). **(c)** Final integrated spectrum. **(d)** Prepared index structure measured by 2-colour interferometry [20] and fitted to even polynomial × exponential.

To identify the important physical effects for ionization injection and acceleration, Fig. 1 presents results from a representative WarpX particle-in-cell (PIC) simulation [27] for the parameters of our experiments, using a 95%/5% $H_2/N_2$ gas mix. The figure shows pulse evolution and electron acceleration for a LWFA drive pulse (blue curve: $a_{0i} = 2.0, \tau_{FWHM} = 35$ fs, $w_0 = 30$ μm Gaussian pulse) injected into a 20 cm waveguide with on-axis plasma density $N_{e0} = 2.0 \times 10^{17}$ cm$^{-3}$, where $w_{ch} = 20$ μm is the $e^{-2}$ intensity radius of the lowest order self-waveguided mode of the structure shown in Fig. 1(d) and computed using our leaky mode solver [9,20,28,29]. The injected pulse is therefore mode-mismatched. Figures 1(a) and 1(b) plot the peak normalized vector potential and electron energy spectrum vs. $z$. The spectrum shows a striated energy structure caused by periodic ionization injection of electrons from $N^{5+}$; by comparing panels it is seen that the individual striations are correlated with the $a_0$ oscillations. With increasing $z$, the striations bend over as dephasing begins, and by $z = 20$ cm the final spectrum of Fig. 1(c) shows multiple quasi-monoenergetic peaks. The characteristic features of these plots are observed for a wide variety of conditions [28]. In pure $H_2$, pulse evolution is similar to that in Fig. 1(a), but with no electron injection and acceleration [28], in agreement with no experiments observing acceleration in pure $H_2$ waveguides.

We identify three stages of pulse evolution as marked on the blue curve in Fig. 1(a): (I) large amplitude intensity oscillations during early propagation, which quickly transition to (II) sustained oscillations at lower amplitude, followed by (III) rapid intensity decline and cessation of oscillations. Periodic ionization injection and the striated bunch energy structure occurs during stages I and II; the degree of dephasing between the electron bunches and the plasma wake during stage III determines whether or not this structure is preserved.

The oscillations in stages I and II are suggestive of beating between different plasma waveguide modes. Although the modes of hydrodynamic plasma waveguides are strictly quasi-bound or "leaky" (owing to the finite thickness cladding or plasma wall) [9,29], an excellent approximation for low order modes of symmetric guides [9, 28] is the spectrum of bound mode wavenumbers of a parabolic plasma profile,

$$\beta_{pm} = k_0 \left[ 1 - \frac{N_{e0}}{2N_{cr}} - \frac{2(2p + m + 1)}{k_0^2 w_{ch}^2} \right] , \qquad (1)$$

where $(p, m)$ identifies the radial and azimuthal indices, $N_{cr}$ is the plasma critical density, $w_{ch} = (\pi r_e \Delta N_e)^{-1/2}$, $\Delta N_e = N_e(w_{ch}) - N_e(0)$, $r_e = e^2/mc^2$ is the classical electron radius, and $k_0 = 2\pi/\lambda_0$ is the vacuum wavenumber [9,28]. The spatial beat period for two interfering modes is $\Lambda = 2\pi/|\beta_{p'm'} - \beta_{pm}|$. Mode beating between the (0,0) and (1,0) plasma waveguide modes has already been measured and simulated in ref. [20]. Each mode has a characteristic $1/e$ propagation decay length, $L_{1/e}$, which is longest for the (0,0) mode and decreases for higher order modes [9]. In the blue curve in Fig. 1(a), the oscillation period in stage I is $\Lambda \sim 4.4$ mm, corresponding to beating dominantly between the (0,0) and (1,0) modes [28], with higher order modes contributing higher frequency interference. Despite the $L_{1/e} \sim 7$ m and $\sim 1$ m decay lengths of the (0,0) and (1,0) modes (both $\gg 20$ cm), the stage I beating quickly transitions to the lower amplitude oscillations of stage II.

To obtain further insight, we examine Fig. 2. Figures 2(a)-(c) show the laser field magnitude $a_0$ in the $xz$ plane; the right column in 2(a) shows the field profile in the transverse ($xy$) plane integrated over the pulse envelope. All plots are normalized to the peak value in the upper left panel of Fig. 2(a). The panels are ordered vertically by the $z$-locations marked on the blue curve in Fig. 1(a). The large amplitude oscillations of stage I manifest in Fig. 2(a) as mode shape and size oscillations corresponding to the (0,0) and (1,0) modes. By $z = 25$ mm, as the stage I beating is fading, a small intensity tail has split off, lagging the main pulse. By $z = 50$ mm, already in stage II, a sequence of evenly spaced high order pulselets has increasingly separated from the main pulse, with the highest order modes the most delayed. This is understood as group velocity walk-off of higher order modes according to $v_{g,pm} = (\partial \beta_{pm}/\partial \omega)^{-1} = c[1 - N_{e0}/2N_{cr} - 2(2p + m + 1)/k_0^2 w_{ch}^2]$ (from Eq. (1) and [9,20]). However, as seen in the panels in Fig 2(b) and the associated propagation movies in [28], the (1,0) structure remains tethered to the (0,0) structure by a long tail, sustaining the beating until $z > \sim 110$ mm, whereupon the stage II oscillations decay as the wake weakens. Owing to the strong $w_{ch}$ dependence, this effect may be less observable in capillary waveguides, which typically have $w_{ch} > 50$ µm.

Examination of additional plots in Fig. 1(a) explains the propagation physics. A low intensity, *mode-mismatched* injected pulse with $a_{0i} = 0.3$ and $w_0 = 30$ µm (orange curve), reproduces the stage I mode-beating of the blue curve; here the beating terminates with complete walk-off of the (1,0) mode. For a higher intensity *mode-matched* Gaussian pulse in a parabolic channel ($a_{0i} = 2.5$, $\tau_{FWHM} = 35$ fs, $w_0 = 30$ µm, $w_{ch} = 30$ µm, and $N_{e0} = 4.0 \times 10^{17}$ cm$^{-3}$ (gold curve)), the stage I oscillations of the blue curve are absent, replaced by regular low amplitude oscillations (seen in [30-33]) extending through $z > 110$ mm (the mode-matched case of $a_0 = 2.0$ and $N_{e0} = 2.0 \times 10^{17}$ cm$^{-3}$ yields little injection and acceleration). These oscillations are similar to the blue curve stage II oscillations, and both can be understood as beating of the (0,0) and (1,0) modes of a *ponderomotively modified channel*, where $w_{ch}^{blue} = 18$ µm (in stage II) and $w_{ch}^{gold} = 24$ µm are determined from the simulations [28]. Each is less than its corresponding unperturbed channel mode radius $w_{ch,0}$ (20µm and 30µm respectively).

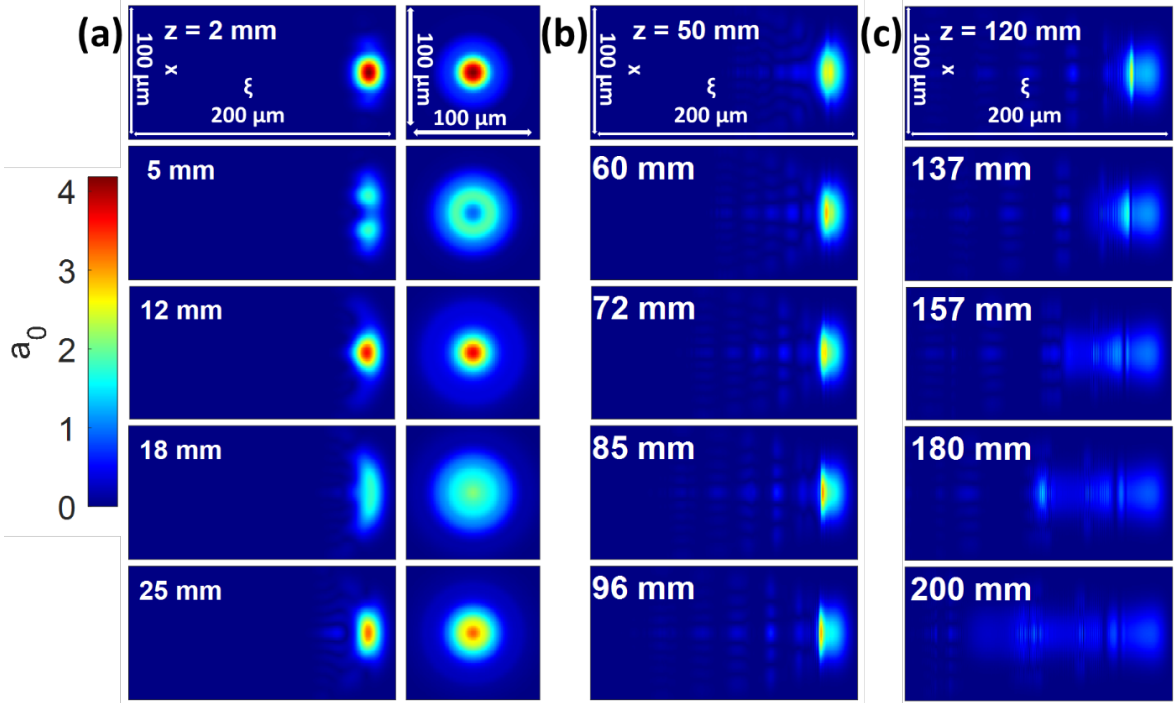

**Figure 2.** Evolution of LWFA drive pulse with $a_{0i} = 2.0, w_0 = 30$ µm, $\tau_{fwhm} = 35\,fs$ in a prepared index structure with $w_{ch} \approx 20\,\mu m$ and on axis plasma density $N_{e0} = 2.0 \times 10^{17}$ cm$^{-3}$. Column **(a)** for stage I, plots $x\xi$ slices of the field magnitude $a_0$ on the left and the $\xi$-integrated mode on the right, where $\xi = z - ct$ is the local space coordinate in the frame of the pulse. Longitudinal positions denoted on the frames are marked by the blue crosses in Fig. 1(a). Column **(b)**, for stage II, plots $x\xi$ slices of $a_0$ for the longitudinal positions marked by the blue stars. Column **(c)**, for stage III, plots $x\xi$ slices of $a_0$ for the longitudinal positions marked by the open circles in Fig. 1(a). All slices and profiles are normalized to the maximum at $z = 2$ mm.

The sustained beating of (0,0) and (1,0) modes appears to be a general feature of $a_0 > 1$ laser pulse propagation in a plasma waveguide, *whether or not the injected pulse is linearly mode-matched to the guide*. As seen in the simulation movies [28], pulse front erosion and red shifting from wake excitation, along with self-steepening at the back of the pulse, act to continuously feed energy from the front of the pulse, where $w_{ch} = w_{ch,0}$, to the cavitated region of the pulse centroid, where $w_{ch} < w_{ch,0}$. This effective mismatch couples to the (0,0) and (1,0) modes, leading to a beat period $\Lambda = 2\pi/|\beta_{10} - \beta_{00}| = \pi^2 \lambda_0 (w_{ch}/\lambda_0)^2$. For the specific cases of Fig. 1(a), $w_{ch} = w_{ch}^{blue}$ (in stage II) gives $\Lambda_{blue} = 4.0$ mm and $w_{ch} = w_{ch}^{gold}$ gives $\Lambda_{gold} = 7.1$ mm. These are in good agreement with the oscillation periods from the simulations in Fig. 1(a).

The striated electron spectrum of Fig. 1(b) shows that the stage II mode beating is sufficient to enhance ionization injection periodically (see Appendix B) from modulation of the wake velocity accompanying laser centroid oscillations [33]. While some injection occurs during stage I, the large amplitude beating induces significant distortion of the wake and loss of trapped electrons. By $z > 80$ mm, ionization injection in stage II ceases owing to pulse self-steepening and etching [28]. By stage III (Fig. 2(c)), the pulse has dispersed and broken up from red shifting and pulse stretching. Here, depletion-induced dephasing [6,34] results in a reduction of maximum accelerated electron energy for $z > 140$ mm, but the striated energy structure is preserved because the pulse is too weak to disrupt the wakefield structure. For higher laser intensity and plasma density, we observe pulse breakup earlier in propagation, leading to wakefield distortion and loss of the striated energy structure.

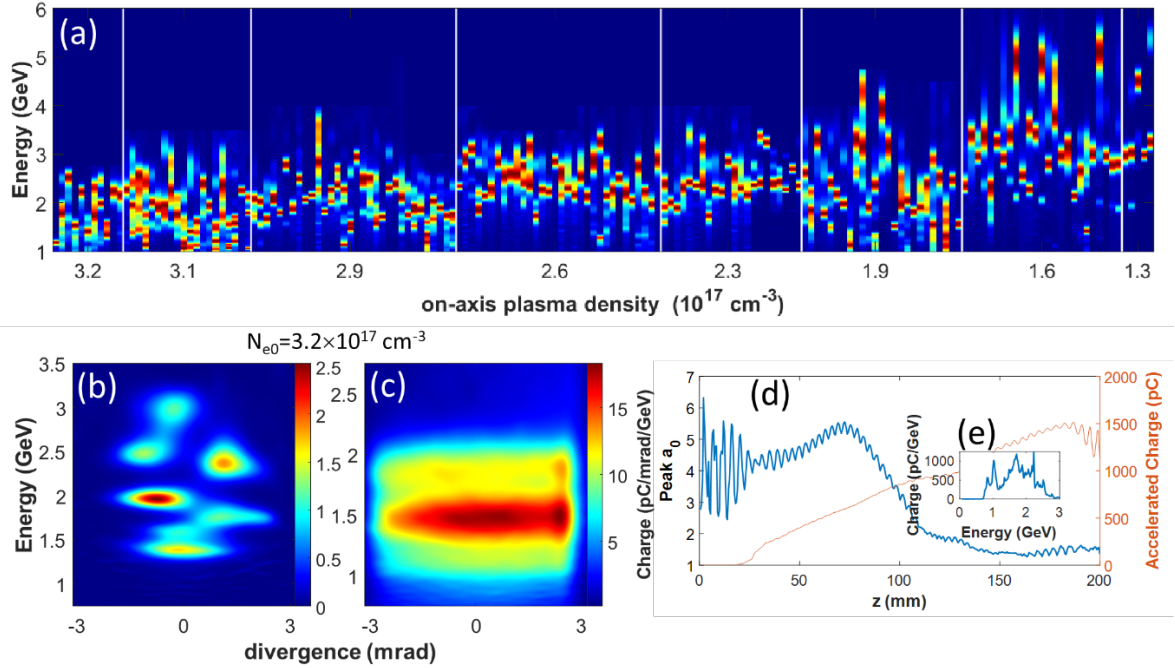

**Figure 3. (a)** Angularly integrated electron spectra varying the gas jet properties (P1 was 11 J, $\tau_{FWHM} = 55$ fs, $w_0 = 30 \mu m$ and $a_0 \sim 2.0$), showing striated multi-peak spectra ranging from $> 1$ GeV to $< 6$ GeV. Here, a 95%/5% $H_2/N_2$ gas mix was injected by all 6 valves along the gas jet. Multiple shots were taken in each density bin (delineated by vertical white lines), with an increase in accelerated electron energy as density drops. Each spectrum is normalized to its peak. **(b)** Example of striated spectrum with $a_{0i} = 2$ and $N_{e0} = 3.2 \times 10^{17}$ cm$^{-3}$. **(c)** Example of 'washed out' spectrum for higher intensity and waveguide density: $a_{0i} = 2.5$ and $N_{e0} = 3.2 \times 10^{17}$ cm$^{-3}$. **(d)** and **(e)**: results of PIC simulation for conditions of (c) showing early onset of depletion. Charge is for $> 1$ GeV.

The three-stage picture is well matched to our LWFA experiments at the ALEPH laser system [35] at Colorado State University. Plasma waveguides were generated by initial Bessel beam heating, via OFI, of a 20 cm long multi-valve-fed gas jet followed by self-waveguiding of the main end-injected pulse [20], which also drives LWFA. A sketch of the experimental setup and a description of the components and parameters are presented in Appendix A. We used three gas flow configurations in our experiments: pure $H_2$ in all valves, a dopant gas mix (95%/5% $H_2/N_2$) in all valves, or pure $H_2$ in all valves with added localized injection of the dopant gas mix.

Figure 3(a) is a plot of angle-integrated electron bunch spectra vs. plasma waveguide central density $N_{e0}$ for the case of the dopant gas mix supplied to all 6 jet valves. Here P1 (11 J, $\tau_{FWHM} = $ 55 fs, $w_0 = 30 \mu m$ and $a_{0i} \sim 2.0$) was injected at 2.5 ns delay into the prepared index structure generated by P2 (0.5 J, $\tau_{FWHM} = 75$ fs). The spectra vary shot to shot, even under nominally similar conditions, owing to injected laser pointing fluctuations [6], but the vast majority of the spectra (such as Fig. 3(b)) are multi-peaked, striated spectra consistent with the simulations of Fig. 1 for the case of dopant gas all along the waveguide. The distribution of the peaks shifts to higher energies with decreasing plasma density, as expected [6]. For higher P1 intensity (15 J, $\tau_{FWHM} = $ 45 fs, $w_0 = 30$ μm and $a_0 \sim 2.6$), the striated spectrum gives way to a very broad, high charge and lower energy peak as illustrated by Fig. 3(c). The simulation of Fig. 3(d) for these conditions shows that higher laser intensity causes an earlier onset of depletion induced dephasing (phase III of pulse propagation), resulting in earlier rollover of the spectrum (compared to Fig. 1(b)), suppressing the lower energy striations and washing out the higher energy ones into a broad peak.

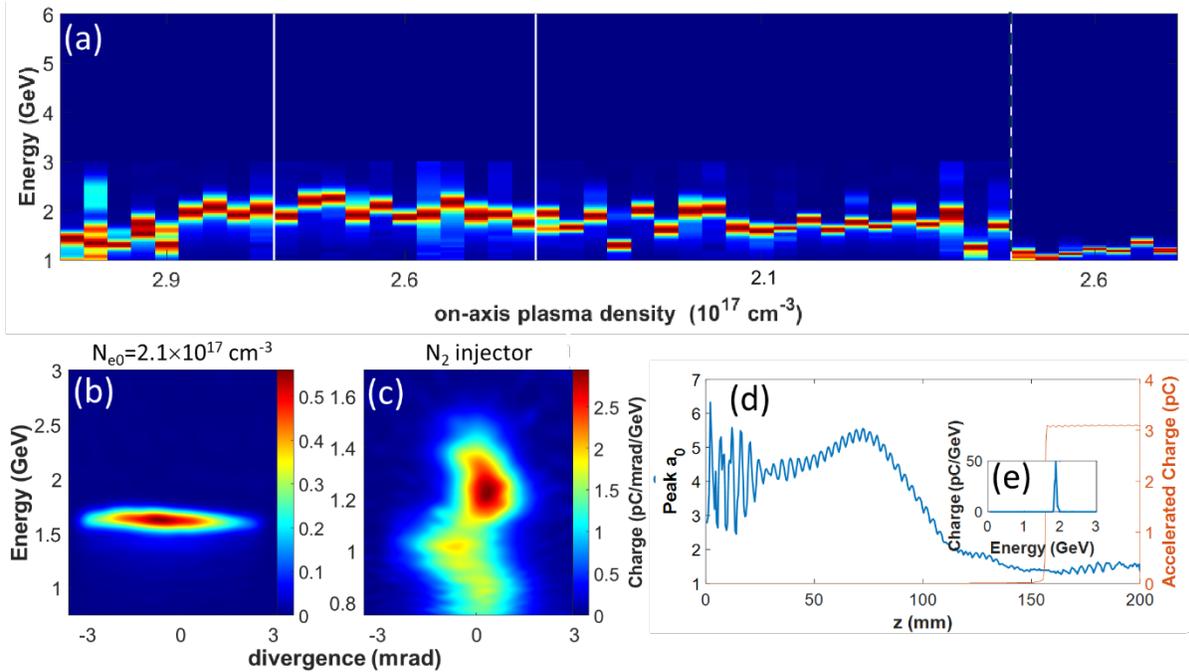

**Figure 4.** (a) Angularly integrated electron spectra from multiple shots in each density bin, showing mostly single quasi-monoenergetic peaks < 2.5 GeV. Here, pure $H_2$ was injected by all 6 valves along the gas jet, and a 95%/5% $H_2/N_2$ dopant gas mix was injected either at $z = 10$ cm (spectra to the left of the dashed white line) or $z = 14$ cm (spectra to the right of the dashed white line, where the waveguide central plasma density was fixed at $N_{e0} = 2.6 \times 10^{17}$ cm$^{-3}$). Each spectrum is normalized to its peak. (b) Example of quasi-monoenergetic spectrum with $a_{0i} = 2.0$ and $N_{e0} = 2.1 \times 10^{17}$ cm$^{-3}$. (c) Example of spectrum resulting from pure $N_2$ injector at or $z = 14$ cm, showing continuous injection. (d) and (e): results of WarpX simulation for conditions of (b). Charge is for $> 1$ GeV.

To generate a quasi-monoenergetic spectrum, these results suggest employing the intensity oscillations as an injector over a localized region of dopant gas. Our gas jet design ([22], Appendix A) provides this capability via axially localized dopant gas inlets. Localized dopant injection has also been used in capillary discharge waveguides to generate single peak electron spectra [36, 37]. Figure 4(a) is a plot of angle-integrated electron bunch spectra vs. plasma waveguide central density $N_{e0}$ for the case of pure $H_2$ supplied to all 6 jet valves, and the dopant gas mix supplied through an injector either at $z = 10$ cm (left side of white dashed line) or at $z = 14$ cm (right side). The injector positions were chosen to be well beyond the stage I mode beating oscillations. Here, as in Fig. 3, the drive pulse P1 was injected at 2.5 ns delay after the Bessel beam pulse P2 (see Appendix A). The most important observation is that the spectra are dominantly single quasi-monoenergetic peaks; they do not have the striated multi-peak structure of Fig. 3. By restricting the dopant gas region, we have limited the number of electron injection events from longitudinal oscillations in pulse intensity. This is reproduced in the simulation (Fig. 4(d) and(e)). The narrowest peaks in Fig. 4 show energy spreads of <10%. Ray tracing simulations [38] of electron beam propagation though our magnetic spectrometer suggest that this number is an overestimate due to additional angular dispersion during transmission through the aluminum chamber exit flange.

To summarize, we have presented experiments and simulations showing that low order mode beating in meter-scale low density plasma waveguides –*occurring whether or not the injected laser is linearly mode matched*– promotes ionization injection to produce multi-peak, multi-GeV electron spectra or single quasi-monoenergetic peaks. Additional spectrum control may be possible using waveguide sections with much higher dopant concentration. For example, Fig. 4(c) shows a spectrum where the dopant gas mix, at $z = 14$ cm, was replaced with pure $N_2$ gas. This produces a continuous spectrum with a peak at ~1.3 GeV in a narrow, ~milliradian angular beam profile. Because the Bessel-beam-induced shock expansion in nitrogen is significantly slower than for hydrogen, the index structure in the injector section is narrowed. The guided pulse is then forced to squeeze into a smaller channel at the injection section, increasing the pulse intensity above the ionization injection threshold and promoting continuous ionization injection. Across all of our experimental conditions, such continuous spectra were generated only with the use of a pure $N_2$ injector, which suggests the possibility of electron spectrum control by guided mode shaping.

The authors thank Mayank Gupta for lab assistance, Scott Wilks for technical discussions, and Revathi Jambunathan and Jean-Luc Vay for help with WarpX [20], the open-source particle-in-cell code used for the simulations of this paper. This work was supported by the U.S. DoE (DE-SC0015516, LaserNetUS DE-SC0019076/FWP#SCW1668, and DE-SC0011375), and NSF (PHY2010511). Simulations used DoD high performance computing support were provided through ONR (N00014-20-1-2233). WarpX (https://github.com/ECP-WarpX/WarpX) is primarily funded by the US DoE Exascale Computing Project, with contributors at LBNL, LLNL, CEA-LIDYL, SLAC, DESY, CERN, and TAE Technologies. J.J.R. acknowledges support of ONR (N000142012842). A.J.G. and A.P. are supported by supported by the Director, Office of Science, Office of High Energy Physics, of the U.S. Department of Energy under Contract No. DE-AC02-05CH11231. E.R. is supported by an NSF Graduate Research Fellowship (DGE 1840340).

*Appendix A: Experimental setup–* Figure 5(a) is a sketch of the experimental configuration. We used three gas flow configurations in our experiments: pure $H_2$ in all valves, a dopant gas mix (95%/5% $H_2/N_2$) in all valves, or pure $H_2$ in all valves with added localized injection of the dopant gas mix. Figure 5(b) plots longitudinal atomic hydrogen gas density profiles [22] used in the experiments. The axial nonuniformity originates from the 6 discrete gas supply valves and slight machining variations in the jet throat. Further details on diagnostics of the gas jet and plasma waveguide and on phase front correction of Bessel beams are found in [6, 22, 39]. The ability of the injector to axially localize the dopant gas is demonstrated in Fig. 5(c). The jet was first run with $H_2$ in the 6 valves and pure $N_2$ in the injector valve, with the fluorescence image collected through an interference filter passing only the H-α emission. The blue curve is the extracted axial density profile of atomic hydrogen, showing a notch at the location of the injector. Then the gases were switched, with $H_2$ in the injector and $N_2$ in the 6 valves, yielding the orange curve showing the axial specificity of injection.

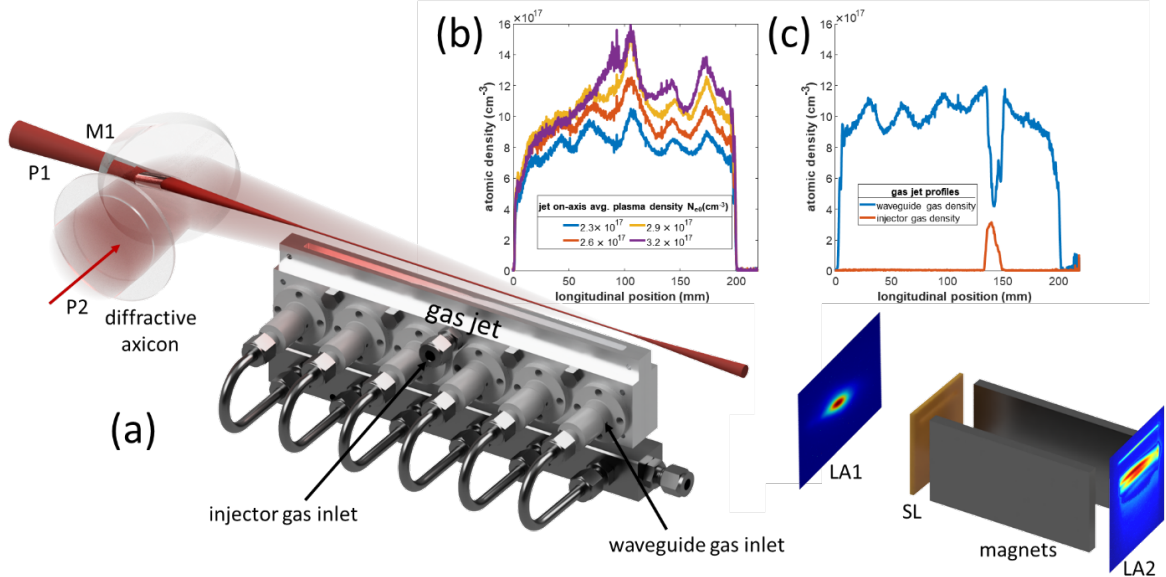

**Figure 5. (a)** Sketch of experiment. A 10-15 J 45-55 fs, $w_0 = 30$ μm drive pulse (P1) is focused into the prepared index structure formed 2.5 ns after an initial plasma column has been generated by a synchronized 0.5 J, 75 fs zeroth-order Bessel beam ($J_0$) pulse (P2) with a ~40 cm long central focal line 3mm above a 20 cm supersonic (Mach 4) $H_2$ gas jet. The $J_0$ pulse is produced by splitting a portion of the main beam, compressing it in a separate compressor, and focusing it with a diffractive axicon. The jet is supplied by 6 gas valves and has localized inlets for dopant gas to seed ionization injection. Accelerated electron bunches exit the end of the waveguide and pass through an 0.85 mm thick aluminum exit flange and LANEX screen (LA1) [40, 41] before entering a 0.93 T permanent magnet spectrometer with an adjustable $0.25 - 1$ mm collimating slit (SL). The dispersed electron spectrum is recorded on LANEX (LA2), and imaged with an Andor Zyla 4.2 camera. **(b)** Atomic hydrogen density profiles and associated on-axis waveguide plasma densities. **(c)** Example of gas distributions with use of a localized $N_2$ injector at $z = 14$ cm from the waveguide entrance.

*Appendix B: Correlation of ionization injection and mode beating–* To investigate the effect of mode beating on electron injection, we performed a quasi-3D simulation with FBPIC [42] using the same laser pulse and plasma waveguide profile as in the blue curve in Fig. 1(a). The simulation was performed in a boosted frame with $\gamma = 5$, with 2 azimuthal modes and 8 particles per cell. The field and tracked particle data are saved every 0.5 mm.

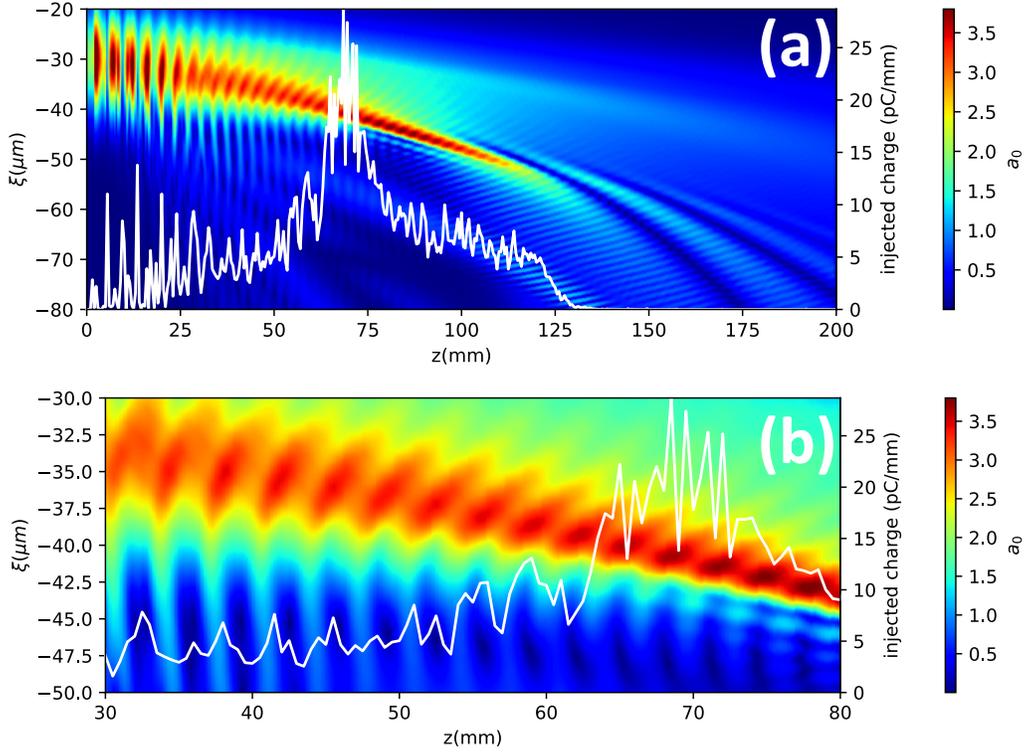

**Figure 6. (a)** Evolution of the axial laser envelope ($a_0$) in the light frame, where $\xi = z - v_g t$ is the local space coordinate and $v_g$ is the laser group velocity along $z$. The white line shows injected charge per unit length over the full propagation. **(b)** Zoomed-in part of panel (a) over $z = 30\sim80\ mm$ showing the correlation between injected charge and the periodic motion of laser pulse peak.

Shown in Fig. 6 are the on-axis electric field envelope in the light frame over the full propagation distance, and injected charge per unit length in each frame. As in Fig. 1(a), we identify the three stages of propagation in panel (a): large amplitude oscillation ($z = 0\sim30$ mm, stage I), sustained lower amplitude oscillations, ($z = 30\sim125$ mm, stage II) and rapid intensity decline ($z > \sim125$ mm, stage III). In stage II, the injected charge is enhanced when the laser peak intensity position retreats in the light frame, which leads to slower wake velocity. Panel (b) shows a zoomed-in part of panel (a) over $z = 30\sim80$ mm. This correlation is clearly seen by the overlap between the peaks of injected charge (white curve) and the periodic deceleration of laser peak position due to mode beating. As a result, the evolution of the electron energy spectra exhibits a striated character. The modulation of laser group velocity due to a mismatched pulse is consistent with simulations in [33].

Downer, *Quasi-Monoenergetic Laser-Plasma Acceleration of Electrons to 2 GeV*, Nat. Commun. **4**, 1988 (2013).
4. H. T. Kim, K. H. Pae, H. J. Cha, I. J. Kim, T. J. Yu, J. H. Sung, S. K. Lee, T. M. Jeong, and J. Lee, *Enhancement of Electron Energy to the Multi-GeV Regime by a Dual-Stage Laser-Wakefield Accelerator Pumped by Petawatt Laser Pulses*, Phys. Rev. Lett. **111**, 165002 (2013).
5. A. J. Gonsalves, K. Nakamura, J. Daniels, C. Benedetti, C. Pieronek, T. C. H. de Raadt, S. Steinke, J. H. Bin, S. S. Bulanov, J. van Tilborg, C. G. R. Geddes, C. B. Schroeder, C. Tóth, E. Esarey, K. Swanson, L. Fan-Chiang, G. Bagdasarov, N. Bobrova, V. Gasilov, G. Korn, P. Sasorov, and W. P. Leemans, *Petawatt Laser Guiding and Electron Beam Acceleration to 8 GeV in a Laser-Heated Capillary Discharge Waveguide*, Phys. Rev. Lett. **122**, 084801 (2019).
6. B. Miao, J.E. Shrock, L. Feder, R.C. Hollinger, J. Morrison, R. Nedbailo, A. Picksley, H. Song, S. Wang, J.J. Rocca, and H.M. Milchberg, *Multi-GeV Electron Bunches from an All-Optical Laser Wakefield Accelerator*, Phys. Rev. X **12**, 31038 (2022).
7. C. G. Durfee and H. M. Milchberg, *Light Pipe for High Intensity Laser Pulses*, Phys. Rev. Lett. **71**, 2409 (1993).
8. C. G. Durfee, J. Lynch, and H. M. Milchberg, *Development of a Plasma Waveguide for High-Intensity Laser Pulses*, Phys. Rev. E **51**, 2368 (1995).
9. T. R. Clark and H. M. Milchberg, *Optical Mode Structure of the Plasma Waveguide*, Phys. Rev. E **61**, 1954 (2000).
10. Y. Ehrlich, C. Cohen, A. Zigler, J. Krall, P. Sprangle, and E. Esarey, *Guiding of High Intensity Laser Pulses in Straight and Curved Plasma Channel Experiments*, Phys. Rev. Lett. **77**, 4186 (1996).
11. A. Butler, D. J. Spence, and S. M. Hooker, *Guiding of High-Intensity Laser Pulses with a Hydrogen-Filled Capillary Discharge Waveguide*, Phys. Rev. Lett. **89**, 185003 (2002).
12. N. Lemos, T. Grismayer, L. Cardoso, G. Figueira, R. Issac, D. A. Jaroszynski, and J. M. Dias, *Plasma expansion into a waveguide created by a linearly polarized femtosecond laser pulse*, Phys. Plasmas 20, 063102 (2013).
13. N. Lemos, L. Cardoso, J. Geada, G. Figueira, F. Albert, and J. M. Dias, *Guiding of Laser Pulses in Plasma Waveguides Created by Linearly-Polarized Femtosecond Laser Pulses*, Sci. Rep. **8**, 3165 (2018).
14. R. J. Shalloo, C. Arran, L. Corner, J. Holloway, J. Jonnerby, R. Walczak, H. M. Milchberg, and S. M. Hooker, *Hydrodynamic optical-field-ionized plasma channels*, Phys. Rev. E 97, 053203 (2018).
15. R. J. Shalloo, C. Arran, A. Picksley, A. von Boetticher, L. Corner, J. Holloway, G. Hine, J. Jonnerby, H. M. Milchberg, C. Thornton, R. Walczak, and S. M. Hooker, *Low-density hydrodynamic optical-field-ionized plasma channels generated with an axicon lens*, Phys. Rev. Accel. Beams 22, 041302 (2019).
16. S. Smartsev, C. Caizergues, K. Oubrerie, J. Gautier, J.-P. Goddet, A. Tafzi, K. T. Phuoc, V. Malka, and C. Thaury, *Axiparabola: A Long-Focal-Depth, High-Resolution Mirror for Broadband High-Intensity Lasers*, Opt. Lett. **44**, 3414 (2019).
17. A. Picksley, A. Alejo, J. Cowley, N. Bourgeois, L. Corner, L. Feder, J. Holloway, H. Jones, J. Jonnerby, H. M. Milchberg, L. R. Reid, A. J. Ross, R. Walczak, and S. M. Hooker, *Guiding of High-Intensity Laser Pulses in 100-mm-Long Hydrodynamic Optical-Field-Ionized Plasma Channels*, Phys. Rev. Accel. Beams **23**, 081303 (2020).

# Supplementary Material: Modal evolution and ionization injection in meter-scale multi-GeV laser wakefield accelerators

J.E. Shrock[1], E. Rockafellow[1], B. Miao[1], M. Le[1], R. C. Hollinger[2], S. Wang[2], A. Picksley[3], J. J. Rocca[2], and H.M. Milchberg[1,4]

[1]Institute for Research in Electronics and Applied Physics and Department of Physics, University of Maryland, College Park, Maryland 20742, USA
[2]Department of Electrical and Computer Engineering, Colorado State University, Fort Collins, Colorado 80523, USA
[3]Lawrence Berkeley National Laboratory, Berkeley, California 94720, USA
[4]Department of Electrical and Computer Engineering, University of Maryland, College Park, Maryland 20742, USA

## 1. Plasma waveguide modes

### (a) Quasi-bound modes

As discussed in Appendix B of ref. [1] and in ref. [2], the quasi-bound modes of a plasma waveguide are found from the function $\eta(\beta') = (|\mathcal{E}_{vacuum}|^2 A)^{-1} \int_A |\mathcal{E}|^2 dA$, where $\mathcal{E}(r)$ is the solution to the Helmholtz equation

$$\frac{d^2\mathcal{E}}{ds^2} + \frac{1}{s}\frac{d\mathcal{E}}{ds} + \left(n^2(s) - \frac{\beta^2}{k_0^2} - \frac{m^2}{s^2}\right)\mathcal{E} = 0 , \qquad (S1)$$

$A$ is the cross-sectional area of the waveguide, and $\mathcal{E}_{vacuum}$ is the amplitude of the vacuum solution. Here, $E(r,z) = \mathcal{E}(r)e^{i\beta z}$ is the quasi-bound propagating electric field mode of the waveguide, $k_0$ is the vacuum wavenumber, $\beta$ and $\beta' = \beta/k_0$ are the unnormalized and normalized longitudinal propagation numbers, $s = k_0 r$ is a dimensionless radial coordinate, $n(s)$ is the radial refractive index profile of the plasma waveguide, and $m$ is an integer azimuthal index.

To find $\eta(\beta')$, Eq. S1 is numerically solved for $\mathcal{E}(s)$ for a range of $\beta'$ for fixed $m$, and the calculated solutions are used to compute $\eta(\beta')$. Sample $\eta(\beta')$ curves for $m = 0$ (blue line) and $m = 1$ (orange line) are shown in Fig. S1. The quasi-bound modes for a given $m$ are located at $\beta'$ values where $\eta(\beta')$ has a local resonance maximum; $\eta(\beta')$ can be viewed as proportional to the frequency dependent Q-factor of a leaky cylindrical resonator. The peaks are ordered as $p = 0,1,2 ...$ for decreasing $\beta'$. All radial modes $p$ associated with a given $m$ can be identified by this process provided that the range of $\beta'$ is large enough and the sampling frequency is sufficient to resolve the resonances. Each peak is also labeled with the $1/e$ attenuation length $L_{1/e}$ of its corresponding $(p,m)$ quasi-bound mode, where $L_{1/e}$ is the FWHM of a Lorentzian fit to the peak [1,2].

The beat frequency between two modes is the difference between their propagation wavenumbers, $|\beta_2 - \beta_1|$, giving a beat period $\Lambda = 2\pi/|\beta_2 - \beta_1|$. Examples of $\Lambda$ are given in the inset table for several low order modes. The plasma profile used for calculating $\eta(\beta')$ and in our particle-in-cell simulations is derived from experimental measurements of plasma and neutral shock expansion using 2-color transverse interferometry [3,4]. The modal analysis assumes full ionization of the core, neutral shock, and background neutrals beyond the shock up to 90 μm away from the axis, where the wings of a $10^{19}$ W/cm² peak intensity Gaussian pulse fall below the OFI threshold for hydrogen.

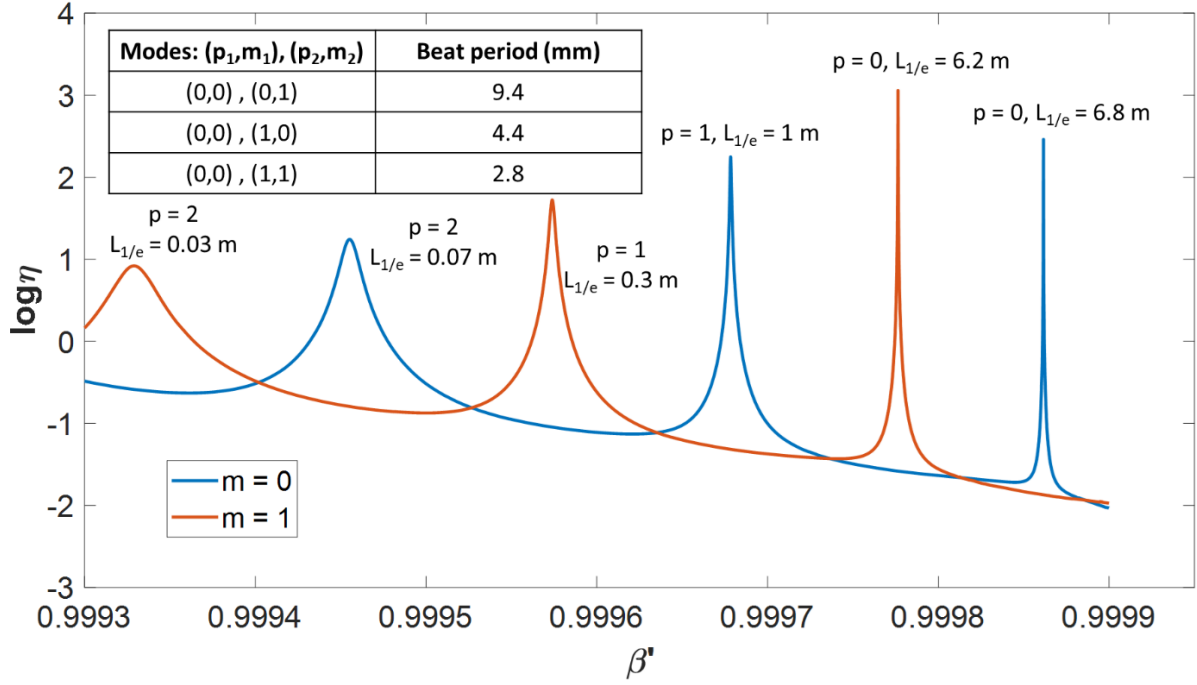

**Figure S1.** Example: $log_{10}\,\eta$ vs. $\beta'$ for a low-density plasma waveguide.

For an infinite parabolic channel, the solutions to Eq. S1 are the Laguerre-Gaussian modes $LG_{pm}$, with wavenumbers $\beta_{pm}$ given by Eq. (1) of the main paper, and lowest order mode spot size $w_0 = w_{ch} = (\pi r_e \Delta N_e)^{-1/2}$. Here $r_e = e^2/m_e c^2$ is the classical electron radius and $\Delta N_e = N(w_{ch}) - N(0)$ [1,2].

**(b) Applicability of quadratic waveguide approximation**

In the main text, we use Eq. (1), derived for bound modes in quadratic index plasma waveguides [2], to approximate the wavenumber for modes of azimuthally symmetric leaky plasma waveguides. This is a very good approximation for a wide range of radial dependence, as we show here. We compute the fundamental mode and wavenumber for a plasma waveguide with the density profile

$$N_e(r) = N_{e0} + \Delta N_e \left(\frac{r}{w_{ch}}\right)^q \quad , q = 2, 3, \ldots 9 , \tag{S2}$$

where $\Delta N_e = N_e(w_{ch}) - N_{e0}$. For $q = 2$, $w_{ch}$ is the exact $1/e$ field radius of the fundamental mode. We take $N_{e0} = 10^{18}$ cm⁻³, $w_{ch} = 30$ μm, and a shock wall thickness of ~20 μm, consistent with measurements in [3]. Figure S2 plots the fundamental mode solutions (($(p, m) = (0,0)$)) and the wavenumbers $\beta_{00}$ and $\beta_{10}$ corresponding to the fundamental mode and first order radial mode, for $q = 2 - 9$. The modal $1/e$ field radius $w_{ch}$ varies by ~10% over this range, $|\beta_{00}/k_0 - 1|$ varies by ~$3 \times 10^{-6}$, and $|\beta_{10}/k_0 - 1|$ varies by ~$5 \times 10^{-5}$, where the terms $N_{e0}/2N_{cr}$ and $2(2p + m + 1)/k_0^2 w_{ch}^2$ from Eq. (1) (in the main paper) are of order ~$2 \times 10^{-4}$.

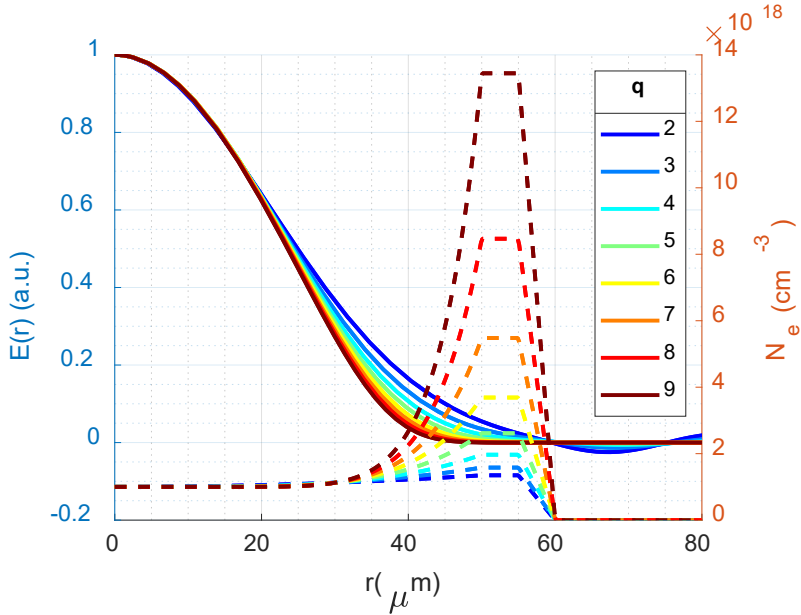

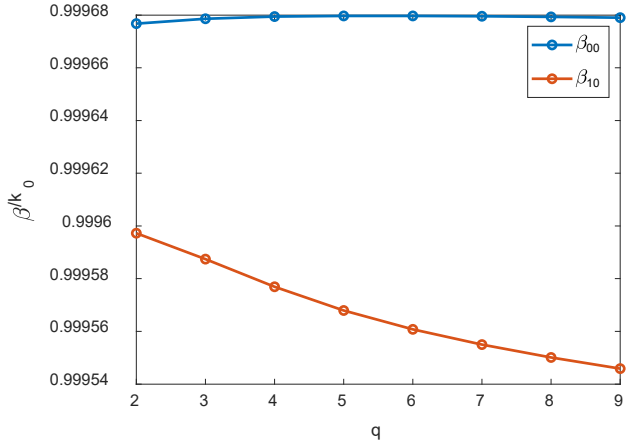

**Figure S2.**

*Top*: Fundamental $(p,m) = (0,0)$ electric field mode profiles $E(r)$ for plasma waveguides with density profile given by $N_e(r) = N_{e0} + \Delta N_e(r/w_{ch})^q$ for $q = 2,\ldots,9$. The field $1/e$ radius $w_{ch}$ varies by 10% over this range of $q$.

*Bottom*: Computed normalized mode wavenumbers $\beta_{00}/k_0$ and $\beta_{10}/k_0$ vs. $q$.

## 2. Additional particle-in-cell simulations

Particle-in-cell simulations were performed for a wide array of laser and plasma conditions using WarpX [5]. All simulations were performed with a $256 \times 256 \times 4096$ grid in the boosted frame corresponding to a lab frame window of $256~\mu m \times 256~\mu m \times 204.8~\mu m$. Figure 1 of the main paper presents a simulation with $a_{0i} = 2.0, w_0 = 30~\mu m, w_{ch} = 20~\mu m, \tau_{fwhm} = 35~\text{fs}, N_{e0} = 2.0 \times 10^{17}~\text{cm}^{-3}$. Results from additional simulations featuring ionization injection and acceleration are plotted in Fig. S3, demonstrating the applicability of our three-stage model under a wide range of laser-plasma conditions. The panels plot the integrated spectrum and $a_0$ vs. propagation distance for the laser-plasma conditions described in the figure caption. Further simulations for parameters resulting in no injection or acceleration are presented in Fig. S4. Panel (a) plots $a_0$ vs. z for mismatched injection of varying intensity pulses with $w_0 = 30~\mu m$, $w_{ch} = 20~\mu m$, and $\tau_{fwhm} = 35~\text{fs}$ in a pure hydrogen plasma waveguide. Additional laser-plasma conditions are included in the legend. The curves in panel (b), for a 95%/5% $H_2/N_2$ doped waveguide, are for varying intensity pulses with $w_0 = 30$ µm, $w_{ch} = 20$ µm, and $\tau_{fwhm} = 35$ fs.

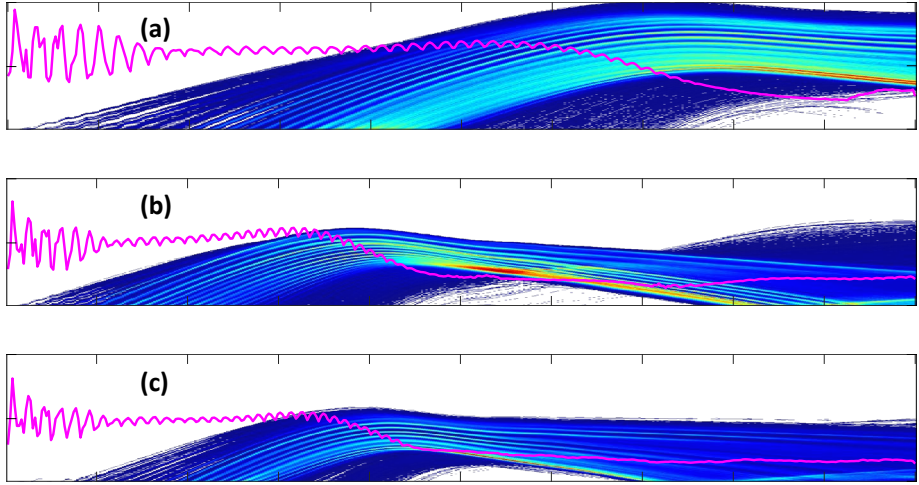

**Figure S3.** Examples of characteristic behavior under different laser-plasma conditions for a plasma waveguide with $w_{ch} = 20 \, \mu m$. (a) $a_{0i} = 1.8, w_0 = 30 \, \mu m, \tau_{fwhm} = 35 \, fs, N_{e0} = 2.0 \times 10^{17} \, cm^{-3}$, (b) $a_{0i} = 1.5, w_0 = 30 \, \mu m, \tau_{fwhm} = 35 \, fs, N_{e0} = 4.0 \times 10^{17} \, cm^{-3}$, and (c) $a_{0i} = 1.5, w_0 = 30 \, \mu m, \tau_{fwhm} = 35 \, fs, N_{e0} = 4.0 \times 10^{17} \, cm^{-3}$ with a 6 µm transverse offset between the centers of the drive pulse and plasma structure.

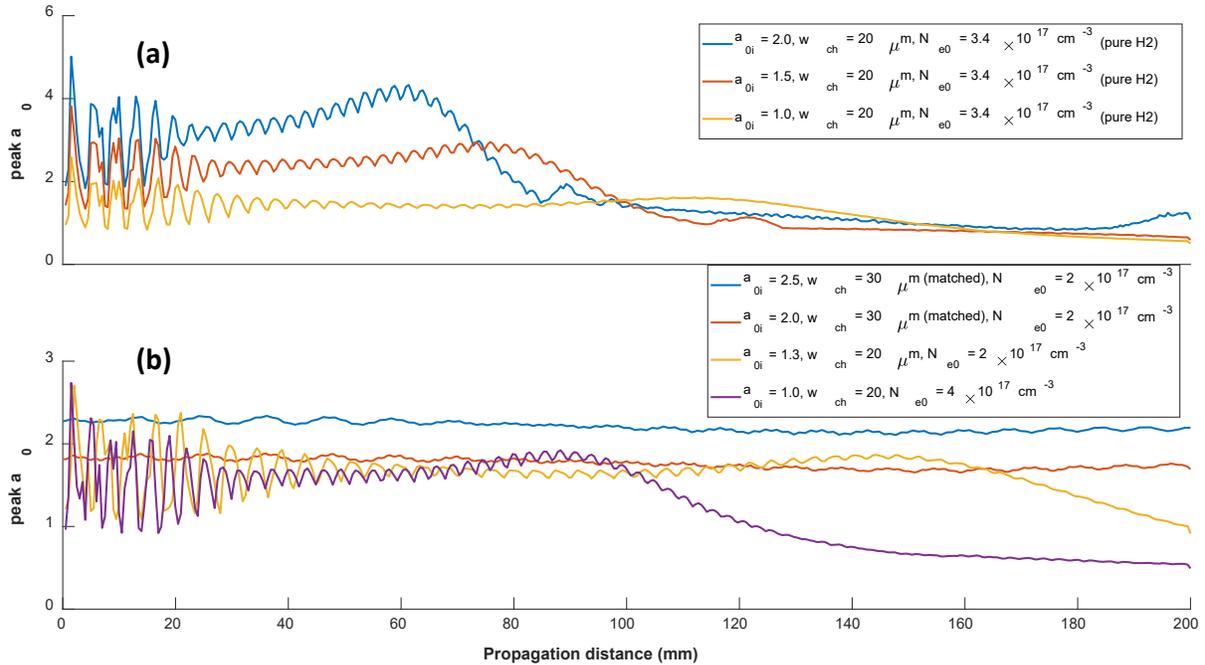

**Figure S4.** Examples of characteristic pulse evolution under various conditions that do not lead to injection and acceleration of electrons. **(a)** peak $a_0$ vs. $z$ for mismatched injection of varying intensity pulses with $w_0 = 30 \, \mu m$, $w_{ch} = 20 \, \mu m$, and $\tau_{fwhm} = 35 \, fs$ in a pure hydrogen plasma waveguide. **(b)** peak $a_0$ vs. $z$ for a 95%/5% $H_2/N_2$ doped waveguide, for varying intensity pulses with $w_0 = 30 \, \mu m$, $w_{ch} = 20 \, \mu m$, and $\tau_{fwhm} = 35 \, fs$. Additional laser-plasma conditions are denoted in the legends of (a) and (b).

## 3. Movies of pulse propagation evolution

Table S1 presents movies of pulse propagation evolution corresponding to the simulation results shown in Fig. 1(a) (and Fig. 2) of the main paper.

**Table S1**

| Conditions | Curve in Fig. 1(a) | Movie links* |
|---|---|---|
| ($i$) mismatched input field $a_{0i} = 2.0, w_0 = 30$ µm, $\tau_{fwhm} = 35\,fs$, $w_{ch} = 20\mu m$, $N_{e0} = 2.0 \times 10^{17}$ cm$^{-3}$ (blue curve) | blue curve | Link ($i$) (a)<br>Link ($i$) (b)<br>Link ($i$) (c)<br>Link ($i$) (d) |
| ($iii$) mismatched input field $a_{0i} = 0.3, w_0 = 30$ µm, $\tau_{fwhm} = 35$ fs, $w_{ch} = 20\mu m$, $N_{e0} = 2.0 \times 10^{17}$ cm$^{-3}$ (orange curve) | orange curve | Link ($iii$) (a)<br>Link ($iii$) (b)<br>Link ($iii$) (c)<br>Link ($iii$) (d) |
| ($iv$) matched input field $a_{0i} = 2.5, w_0 = 30$ µm, $\tau_{fwhm} = 35\,fs$, $w_{ch} = 30\,\mu m$, $N_{e0} = 4.0 \times 10^{17}$ cm$^{-3}$ (gold curve) | gold curve | Link ($iv$) (a)<br>Link ($iv$) (b)<br>Link ($iv$) (c)<br>Link ($iv$) (d) |
| * (a): $a_0(x, \xi; z)$, each frame normalized to $a_0$ maximum in that frame<br>(b): $a_0(x, \xi; z)$, each frame normalized to $a_0$ maximum in full sequence<br>(c): $\log_{10}(a_0(x, \xi; z))$, where $a_0(x, \xi; z)$ is from movie (b)<br>(d): Corresponding $k_z$ spectrum and envelope $a_0(x = 0, \xi; z)$ vs. $z$<br>The pulse local spatial coordinate is $\xi = z - ct$ | | |

## 4. Simple model for beating of the $(p, m) = (0, 0)$ and $(1, 0)$ modes in ponderomotively modified waveguides

As discussed in the main text, beating occurs between the (0,0) and (1,0) modes of the ponderomotively modified waveguide profile. The plot in Fig. S5 is a frame at $z = 80$ mm from the movie ("Link ($iv$)(d)" in Table S1) corresponding to the gold curve in Fig. 1(a) of the main paper. The pulse propagates left to right. It is seen that the spectral red shifting terminates at the location of the self-steepening-induced optical shock at the rear, indicating that the pulse energy centroid is located in a nearly cavitated region.

A simple model of the ponderomotive channel modification provides physical insight. In the long pulse limit, the ponderomotive channel modification is expressed as

$$\frac{N_e}{N_{ei}} = 1 + k_p^{-2}\nabla_\perp^2 \gamma, \qquad (S3)$$

where the relativistic factor is $\gamma = (1 + a^2/2)^{1/2}$ for linearly polarized pulses, and $N_e, N_{ei}$, and $k_p$ are functions of $r$. Here $N_{ei}(r) = N_{e0} + r^2/\pi r_e w_{ch}^4$ [2] is the parabolic plasma density profile of the unperturbed channel, $k_p = c^{-1}(4\pi N_{ei} e^2/m)^{1/2}$ is the plasma wavenumber, and $a = a_{0i}\exp(-(r/w_{ch})^2)$ is the lowest-order mode.

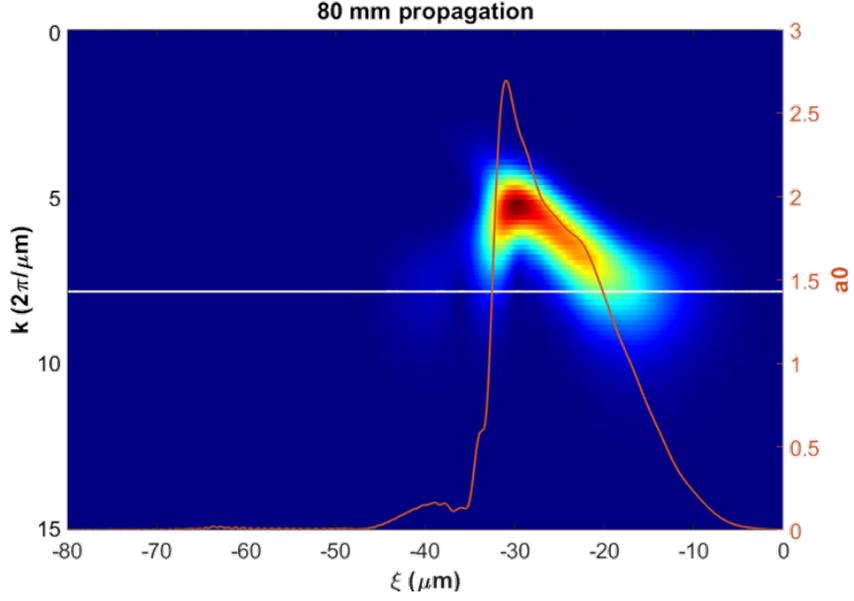

**Figure S5.** Frame at $z = 200\ mm$ from movie "Link $(iii)(d)$", corresponding to conditions $(iii)$ from Table S1.

For the gold curve in Fig. 1(a) of the main paper, the channel modification $\Delta N_e(r) = N_e(r) - N_{ei}(r)$ determined by Eq. (S3) is plotted in Fig. S6. The modification is negligible for $r > 2w_{ch} = 60$ μm.

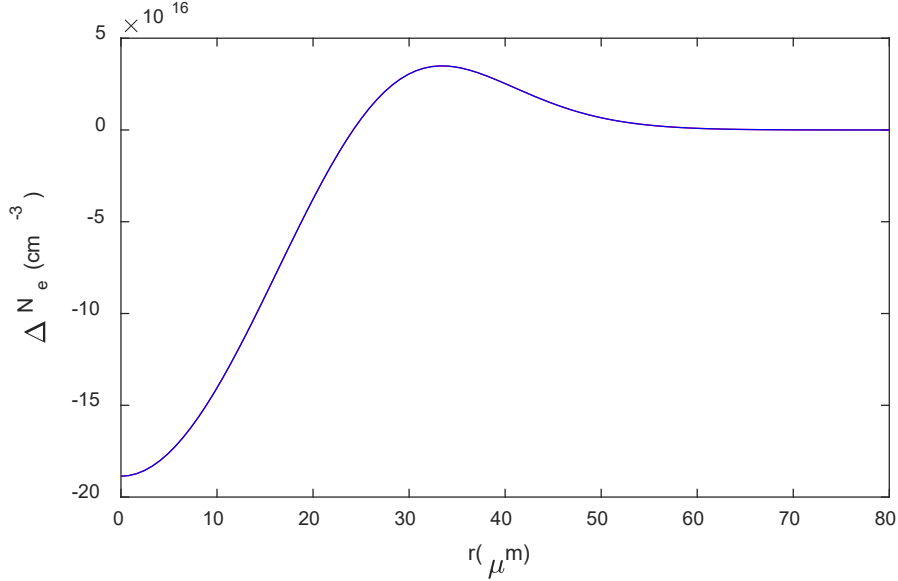

**Figure S6.** Channel modification $\Delta N_e(r) = N_e(r) - N_{ei}(r)$ calculated using Eq. (S2).

Assuming that the ponderomotively perturbed channel is locally parabolic with modified central density $N'_{e0}$ and modified mode radius $w'_{ch}$, we write

$$N_e(r) = N'_{e0} + \frac{r^2}{\pi r_e w'^4_{ch}} \tag{S4}$$

Using $N'_{e0} = 0$ (owing to near-cavitation at the pulse energy centroid) and imposing charge conservation ($\int_0^{2w_{ch}} rdrN_e(r) = \int_0^{2w_{ch}} rdrN_{ei}(r)$) yields an expression for $w'_{ch}$,

$$\frac{1}{w'^4_{ch}} = \frac{1}{w^4_{ch}} + \frac{\pi r_e N_{e0}}{2 w^2_{ch}} \quad , \tag{S5}$$

that can be compared to the local mode radius measured from the WarpX [5] simulations. Table

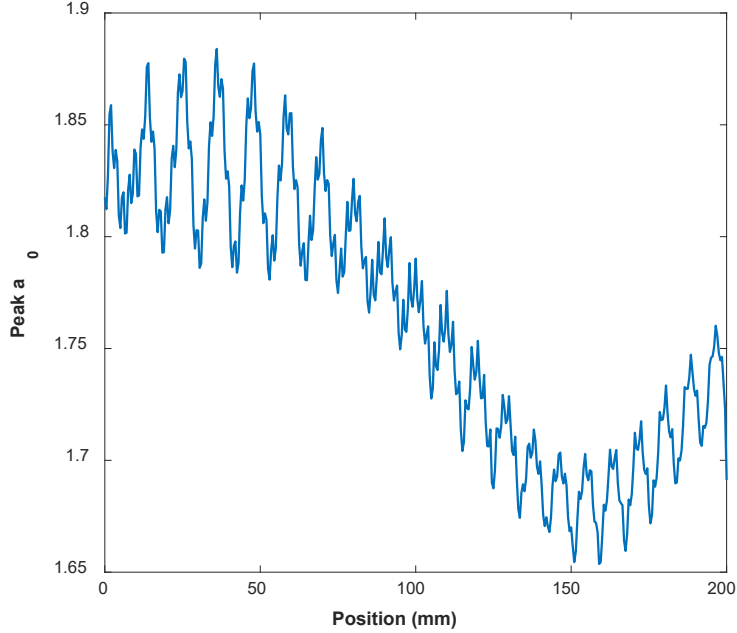

**Figure S7.** Peak $a_0$ vs. z for matched pulse case: $a_{0i} = 2.0$, $w_0 = 30 \ \mu m$, $w_{ch} = 30 \ \mu m$, $N_{e0} = 2.0 \times 10^{17} cm^{-3}$

S2 shows this comparison for three cases, two corresponding to the blue and gold curves of Fig. 1(a) of the main paper and one corresponding to Fig. S7. In all cases, $w'_{ch}$ determined by the simple model agrees well with the value extracted from the PIC simulations, giving confidence in our physical picture of mode structure in ponderomotively modified plasma waveguides. The beat period is calculated as $\Lambda = 2\pi/|\beta_{10} - \beta_{00}|$, where $\beta_{pm}$ is the quasi-bound mode wavenumber approximated by Eq. (1) of the main paper. This gives $\Lambda = \pi^2 \lambda_0 (w'_{ch}/\lambda_0)^2$, with values plotted in column H of Table S2. These values compare well to the values (column G) determined by a continuous wavelet transform (CWT) of the associated simulation curves. We note that little energy is coupled into the (0,1) and higher order azimuthal modes of the ponderomotively modified channel since it is necessarily co-axial with the drive pulse; the process resulting in $(0,0) - (1,0)$ mode beating essentially selects for radial modes.

**Table S2**

| A | B | C | D | E | F | G | H |
|---|---|---|---|---|---|---|---|
| $w_0$, unperturbed $w_{ch}$ ($\mu m$) | $a_{0i}$ | $N_{e0}$ ($10^{17}$ $cm^{-3}$) | $w'_{ch}$ ($\mu m$) from WarpX simulations | $w'_{ch}$ ($\mu m$) from Eq. (S4) | associated simulation curve | $\Lambda$ (mm) from CWT of curve | $\Lambda$ (mm) from $w'_{ch}$ (column D) |
| 30, 20 mismatched | 2.0 | 2.0 | 18 | 18.5 | blue curve, Fig. 1(a) | 3.5 | 4.0 |
| 30, 30 matched | 2.5 | 4.0 | 24 | 23.6 | gold curve, Fig. 1(a) | 6.5 | 7.1 |
| 30, 30 matched | 2.0 | 2.0 | 27 | 25.9 | Fig. S7 | 8.6 | 8.3 |

An example of the relative positions of the different pieces of the pulse with respect to the wake is shown in Fig. S8, which plots the laser pulse envelope overlaid on the plasma density over a full cycle of phase II beating from the blue curve in Fig. 1(a) ($a_0 = 2.0, w_0 = 30$ μm, $\tau_{fwhm} = 35\ fs$, $w_{ch} = 20\mu m$, $N_{e0} = 2.0 \times 10^{17}$ cm$^{-3}$). The complete separation of the higher order modes is clearly seen, as is the sustained beating between the (0,0) and (1,0) components of the pulse, where the (1,0) mode appears in x – ξ cross section as the 3 peaks immediately behind the (0,0) energy centroid. This beating is also clear from the movies at Link$(i)$(a)–Link$(i)$(c) in Table S1.

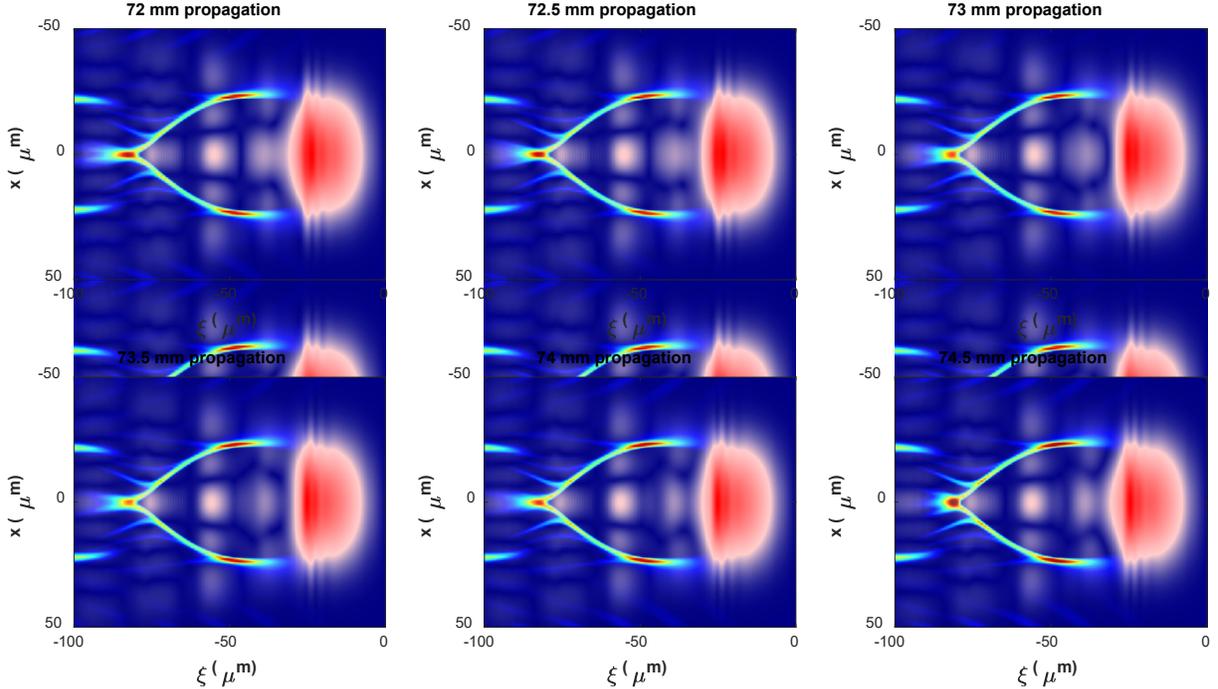

**Figure S8.** Pulse envelope and wake plasma density for a cycle of stage II intensity modulation. The complete separation of the higher order modes is clearly seen, as is a full cycle of beating between the (0,0) and (1,0) components of the pulse, where the (1,0) mode appears in $x - \xi$ cross section as the 3 vertically spaced peaks immediately behind the (0,0) energy centroid.